\documentclass[11pt]{article}
\usepackage{epsfig,amssymb}
\usepackage{graphicx}

\begin{document}

\title{Synchronization of networks of oscillators with distributed delay coupling}

\author{Y.N. Kyrychko\thanks{Corresponding author. Email: y.kyrychko@sussex.ac.uk}, \hspace{0.5cm}K.B. Blyuss 
\\\\ Department of Mathematics, University of Sussex, Falmer,\\
Brighton, BN1 9QH, United Kingdom\\\\
\and E. Sch\"oll
\\\\ Institut f\"ur Theoretische Physik, Technische Universit\"at Berlin,\\
10623 Berlin, Germany}

\maketitle

\begin{abstract}
This paper studies the stability of synchronized states in networks where couplings between nodes are characterized by 
some distributed time delay, and develops a generalized master stability function approach. Using a generic example of Stuart-Landau
oscillators, it is shown how the stability of synchronized solutions in networks with distributed delay coupling can be determined
through a semi-analytic computation of Floquet exponents. The analysis of stability of fully synchronized and of cluster or splay states
is illustrated for several practically important choices of delay distributions and network topologies.

\end{abstract}

\section{Introduction}

In the last decade there has been a substantial growth of research interest in the dynamics of coupled systems, ranging from a few elements to large networks \cite{AB,BB,New,WS}. From a perspective of potential applications, one of the central research questions for such systems is the emergence and stability of different types of collective dynamics and, in particular, synchronization \cite{PRK01,BOC06}. In order to study stability of synchronization, Pecora and Carroll \cite{PC98} put forward a {\it master stability function} (MSF) approach that allows one to separate the local dynamics of individual nodes from the network topology, which is achieved by a proper diagonalization of the matrix representing the full network dynamics.

An important aspect of network dynamics concerns the fact that interactions between nodes are often non-instantaneous due to a finite speed of signal propagation. In order to account for this feature, one has to explicitly include time delays in the analysis, and this is known to have a significant impact on network behaviour \cite{DVDF10, DVEDF08, ES, EHKK11, EKA10, FYDS10, HHFS11, K09, LDHS,ATA10,ZAK13a,FLU13,SOR13}. Some work has been done on the extension of the master stability function to networks with time-delayed coupling \cite{CHO09,CHO11,CHO13,DJD04,K09,DAH12,WIL13} but so far it has only included single constant time delays. Hunt {\it et al.} \cite{Hunt1,Hunt2} have considered complex networks of linearly coupled nodes with time delays and noise and established conditions for synchronizability of such networks. At the same time, in many realistic systems, time delays may be not constant and may either vary depending on the values of system variables, or just not be explicitly known \cite{GU1,GU2,GJU14}, and in these cases the standard methodology of considering one or several constant time delays is not sufficient. To describe such situations mathematically, one one can use the formalism of distributed time delays, where the time delay is represented through an integral kernel describing a particular delay distribution \cite{ATA03,CJ09,KK11,CHO14}. Distributed time delays have been successfully used to describe situations when only an approximate value of time delay is known in engineering experiments \cite{KK11,MVAN05,GJU13}, for modelling distributions of waiting times in epidemiological models \cite{BK10}, maturation periods in population and ecological models \cite{FT10,GS03}, as well as in models of traffic dynamics \cite{SAN07} and neural systems \cite{ETF05,VIC08,WIL14}. In a recent paper, Morarescu {\it et al.} \cite{Mor13} have looked at systems with gamma-distributed delay coupling and analyzed the stability of the synchronized equilibria (steady states), which are not affected by the coupling.

In this paper we consider networks of coupled identical dynamical systems with linear distributed delay coupling as represented by some integral kernel. Linearization near the synchronization manifold yields a variational equation, which after block-diagonalization of the coupling matrix reduces to a single complex-valued integro-differential equation, whose maximum Floquet or Lyapunov exponent gives the master stability function in terms of the system parameters. Using the example of a network of Stuart-Landau oscillators, we make further analytical progress by showing how the computation of the master stability function can be reduced to finding solutions of a single transcendental characteristic equation. For particular choices of the delay distribution kernel, all terms in this equation can be found in a closed analytical form in terms of coupling and system parameters. The methodology we develop is quite generic and can be applied to analyze the stability of fully synchronized and of cluster states in any systems with distributed delay coupling.

The outline of this paper is as follows. In the next section we develop a general master stability function formalism for networks of identical coupled elements with distributed delay coupling. Section 3 illustrates how this general framework can be made more explicit by means of an amplitude-phase representation for the example of coupled Stuart-Landau oscillators. Section 4 contains the results of analytical and numerical computations of the master stability function for coupled Stuart-Landau oscillators with uniform and gamma distributed delays and different coupling topologies. In section 5 the methodology of the master stability function for systems with distributed delay coupling is extended to the analysis of cluster states representing generalized synchronization. The paper concludes with discussion of results in section 6.

\section{Master stability function}

We consider a network of $N$ identical dynamical systems with distributed delay coupling
\begin{equation}\label{eq1}
\dot{{\bf x}}_k={\bf F}_k[{\bf x}_k(t)]+\sigma \sum_{j=1}^{N}G_{kj}{\bf H}_{kj}\left[\int_{0}^{\infty}g(t'){\bf x}_{j}(t-t')dt'-{\bf x}_{k}(t)\right],\hspace{0.5cm}k=1,\ldots, N,
\end{equation}
where ${\bf x}_{k}\in\mathbb{R}^{m}$ is the state vector of each system, ${\bf F}_k[({\bf x}_k(t)]$ describes the local dynamics of the node $k$, $\sigma\in\mathbb{C}$ is the coupling strength, ${\bf G}$ is an $N\times N$ matrix describing the coupling topology and the strength of each link in the network, ${\bf H}_{kj}$ is an $m\times m$ matrix representing the coupling scheme, and $g(\cdot)$ is a delay distribution kernel, satisfying
\[
g(u)\geq 0,\hspace{0.5cm} \int_{0}^{\infty}g(u)du=1.
\]
For $g(u)=\delta(u)$, one recovers instantaneous coupling $({\bf x}_{j}-{\bf x}_{k})$; for $g(u)=\delta(u-\tau)$, the coupling takes the form of a discrete time delay: $[{\bf x}_{j}(t-\tau)-{\bf x}_{k}(t)]$. The matrix $G$, whose non-zero entries $G_{ij}$ define a relative strength of a link between nodes $j$ and $i$, is related to the {\it adjacency matrix} of the network topology, whose elements are zero or one depending on the existence of the respective link.

Next, we discuss a number of assumptions regarding the individual dynamical systems and their interactions in the network. All nodes are assumed to be identical, so that their dynamics is described by a single function ${\bf F}_k[\cdot]={\bf F}[\cdot]$. The time delays in the propagation of signals from different nodes are also taken to be the same and described by the delay distribution kernel $g(u)$. Since we are interested in synchronization and its stability, we will further assume that the matrix ${\bf G}$ has a constant row sum
\begin{equation}\label{row_sum}
\sum_{j=1}^{N}G_{kj}=\mu.
\end{equation}
This condition is necessary for the existence of the synchronized solution. Finally, we assume that the coupling schemes are identical for different nodes, i.e. ${\bf H}_{kj}={\bf H}$ for $k,j=1,\ldots N$. The last two assumptions ensure that all nodes received the same input. With those assumptions, the model (\ref{eq1}) simplifies to
\begin{equation}\label{eq2}
\dot{{\bf x}}_k={\bf F}[{\bf x}_k(t)]+\sigma \sum_{j=1}^{N}G_{kj}{\bf H}\left[\int_{0}^{\infty}g(t'){\bf x}_{j}(t-t')dt'-{\bf x}_{k}(t)\right],\hspace{0.5cm}k=1,\ldots, N.
\end{equation}

Let us now consider the {\it synchronization manifold}, defined as ${\bf x}_k(t)\equiv {\bf x}_s(t)$ for all $k=1,\ldots N$. The dynamics on the synchronization manifold is given by
\begin{equation}\label{sync}
\dot{\bf x}_s={\bf F}[{\bf x}_s(t)]+\sigma \mu{\bf H}\left[\int_{0}^{\infty}g(t'){\bf x}_{s}(t-t')dt'-{\bf x}_{s}(t)\right],
\end{equation}
where $\mu$ is the row sum, Eq.~(\ref{row_sum}). It is worth noting that the dynamics of synchronization manifold is independent of the coupling strength in the case of zero row sum $\mu=0$ or instantaneous coupling $g(u)=\delta(u)$, and in all other cases it will also depend on the coupling strength $\sigma$.

To study the stability of the synchronization manifold, we linearize the full system (\ref{eq2}) near ${\bf x}_k(t)={\bf x}_s(t)$, which yields with ${\bf x}_k(t)={\bf x}_s(t)+\xi_k(t)$,
\begin{equation}
\dot{\xi}_k(t)={\bf J}_0(t)\xi_k(t)+\sigma \sum_{j=1}^{N}G_{kj}{\bf H}\left[\int_{0}^{\infty}g(t')\xi_{j}(t-t')dt'-\xi_{k}(t)\right],\hspace{0.5cm}k=1,\ldots, N,
\end{equation}
where
\[
{\bf J}_0(t)=D{\bf F}[{\bf x}_s(t)]
\]
is the Jacobian of ${\bf F}$. Introducing a vector $\mbox{\boldmath$\xi$}=(\xi_1,\xi_2,\dots,\xi_N)^{T}$, the above equation can be rewritten as
\begin{equation}
\dot{\mbox{\boldmath$\xi$}}={\bf I}_N\otimes [{\bf J}_0(t)-\sigma \mu{\bf H}]\mbox{\boldmath$\xi$}+\sigma [{\bf G}\otimes {\bf H}]\int_{0}^{\infty}g(t')\mbox{\boldmath$\xi$}(t-t')dt',
\end{equation}
where $\otimes$ denotes the Kronecker product, and ${\bf I}_N$ is the $N\times N$ unity matrix. We will assume that matrix
${\bf G}$ is {\it diagonalizable}, i.e., there exists a unitary transformation ${\bf U}$ such that
\[
{\bf UGU}^{-1}={\rm diag}(\mu,\nu_1,\nu_2,\ldots,\nu_{N-1}).
\]
Here the first eigenvalue of the matrix ${\bf G}$, which is equal to the row sum $\nu_0\equiv \mu$, corresponds to the eigenvector $(1,1,\ldots,1)$ describing the dynamics on the synchronization manifold. Hence, this eigenvalue is called the {\it longitudinal eigenvalue} of ${\bf G}$, whereas all other eigenvalues are called {\it transverse eigenvalues} as they correspond to directions transverse to the synchronization manifold.

To make further analytical progress, it is instructive to diagonalize the coupling matrix ${\bf G}$, which results in a block-diagonalized variational equation
\begin{equation}\label{VarEq}
\dot{\zeta}_{k}(t)=\left[{\bf J}_0(t)-\sigma \mu{\bf H}\right]\zeta_k(t)+\sigma \nu_k {\bf H}\int_{0}^{\infty}g(t')\zeta_k(t-t')dt',\hspace{0.5cm}k=0,\ldots, N-1,
\end{equation}
where $\nu_k$ is the $k$-th eigenvalue of the coupling matrix ${\bf G}$. It is worth noting that in this system of equations, the Jacobian ${\bf J}_0(t)$ and ${\bf H}$ are the same for all $k$, and one can consider the variational equation in dependence on the complex parameter $\psi+i\beta$
\begin{equation}\label{MSF_eq}
\dot{\zeta}(t)=\left[{\bf J}_0(t)-\sigma \mu{\bf H}\right]\zeta(t)+(\psi+i\beta) {\bf H}\int_{0}^{\infty}g(t')\zeta(t-t')dt'.
\end{equation}
Computation of the maximum Lyapunov exponent (or real part of the Floquet exponent) for this equation yields the {\it master stability function}
\[
{\rm Re}[\Lambda_{\rm max}(\psi,\beta,\sigma)],
\]
and the analysis of stability of the synchronization manifold reduces to checking that for all transverse eigenvalues of the coupling matrix $\sigma\nu_k=\psi+i\beta$, $k=1,\ldots, N-1$, one has Re[$\Lambda_{\rm max}(\psi,\beta,\sigma)]<0$. A positive value of the Lyapunov exponent (real part of the Floquet exponent) associated with the longitudinal eigenvalue $\nu_0$ indicates that the synchronized solution is chaotic (unstable). The significant advantage of this approach lies in the separation of the actual dynamics from the topology of the coupling, as the computation of $\Lambda_{\rm max}(\psi,\beta,\sigma)$ can be done once for the equation (\ref{MSF_eq}) independently of the coupling matrix. It is worth noting that although the computation of the master stability function can be performed independently of the coupling topology, it has to be done separately for each value of the coupling strength $\sigma$. The reason for this is the fact that the row sum $\mu$ is, in general, non-zero, and also the synchronized state ${\bf x}_s(t)$ itself depends on the coupling strength due to the delayed nature of the coupling, as follows from Eq.~(\ref{sync}). The above approach can be generalized in a straightforward manner to more complex forms of the coupling function ${\bf H}$.

\section{Coupled Stuart-Landau oscillators}

In order to illustrate the analysis of the stability of synchronization, we consider an array of $N$ identical diffusively coupled Stuart-Landau oscillators ($k=1,\ldots, N$)
\begin{equation}\label{SLeq}
\displaystyle{\dot{z}_k(t)=(\lambda+i\omega)z_{k}(t)-(1+i\gamma)|z_k(t)|^2 z_k(t)+\sigma\sum_{j=1}^{N}G_{kj}\left[\int_{0}^{\infty}g(t')z_{j}(t-t')dt'-z_k(t)\right],}
\end{equation}
which is a prototype of dynamics near a supercritical Hopf bifurcation. Here, $z_k\in\mathbb{C}$, $\lambda$, $\omega\neq 0$ and $\gamma$ are real constants, where $\omega$ is the intrinsic oscillation frequency, and
\[
\sigma=Ke^{i\theta},\hspace{0.5cm}K\in\mathbb{R}_{+},\hspace{0.3cm}\theta\in\mathbb{R},
\]
where $K$ and $\theta$ are the strength and the phase of coupling, respectively. Such complex-valued couplings, which are equivalent to a rotational matrix ${\bf H}$ if $z_k$ is written in terms of real and imaginary part, have been shown to be important in overcoming the odd-number limitation of time-delayed feedback control \cite{FFGHS}, in controlling amplitude death \cite{KBS11,KBS13}, and in anticipating chaos synchronization \cite{PP}.

Introducing amplitude and phase variables as $r_k=|z_k|$ and $\phi_k=\arg(z_k)$, the equation (\ref{SLeq}) can be recast as a system of $2N$ real equations
\begin{equation}\label{AmpPhase}
\begin{array}{l}
\displaystyle{\dot{r}_k=[\lambda-r_k^2(t)]r_k(t)
+K\sum_{j=1}^{N}G_{kj}\left[\int_{0}^{\infty}g(t')r_{j}(t-t')\cos[\phi_j(t-t')-\phi_k(t)+\theta ]dt'-r_k(t)\cos\theta\right]},\\
\displaystyle{\dot{\phi}_k=\omega-\gamma r_k^2+K\sum_{j=1}^{N}G_{kj}\left[\int_{0}^{\infty}g(t')\frac{r_{j}(t-t')}{r_k(t)}\sin[\phi_j(t-t')-\phi_k(t)+\theta ]dt'-\sin\theta\right],\hspace{0.1cm}k=1,\ldots, N.}
\end{array}
\end{equation}
The fully synchronized solution of the system (\ref{AmpPhase}) is given by
\begin{equation}\label{SyncSol}
r_k(t)=r_0,\hspace{0.5cm}\phi_{k}=\Omega t,\hspace{0.3cm}k=1,\ldots, N,
\end{equation}
with the common radius $r_0$ and the common frequency $\Omega$ of oscillations being determined by the solutions of the following equations
\begin{equation}\label{r0eq}
\begin{array}{l}
r_{0}^{2}=\lambda+K\mu\left[F_c(-\Omega,\theta)-\cos\theta\right],\\\\
\Omega=\omega-\gamma r_0^2+K\mu\left[F_s(-\Omega,\theta)-\sin\theta\right],
\end{array}
\end{equation}
where we have introduced the auxiliary quantities
\begin{equation}\label{aux_quant}
F_c(a,b)=\int_{0}^{\infty}g(t')\cos(at'+b)dt',\hspace{0.5cm}F_s(a,b)=\int_{0}^{\infty}g(t')\sin(at'+b)dt'.
\end{equation}
To analyse the stability of the fully synchronized solution (\ref{SyncSol}), we use a slightly modified ansatz for small perturbations around this solution, namely,
\[
r_k(t)=r_0[1+\delta r_k(t)],\hspace{0.5cm}\phi_k(t)=\Omega t+\delta \phi_k(t),\hspace{0.5cm}k=1,\ldots,N,
\]
which yields a variational equation of the form
\begin{equation}
\dot{\xi}_k=({\bf J}_0-K\mu{\bf R}_0)\xi_k+K\sum_{j=1}^{N}G_{kj}\int_{0}^{\infty}g(t'){\bf R}_1(t')\xi_j(t-t')dt',\hspace{0.5cm}k=1,\ldots,N,
\end{equation}
where $\xi_k=(\delta r_k,\delta \phi_k)^{T}$, and the matrices ${\bf J}_0$, ${\bf R}_0$ and ${\bf R}_1$ are given by
\[
{\bf J}_0=\left(
\begin{array}{cc}
-2r_0^2&0\\
-2\gamma r_0^2&0
\end{array}
\right),
{\bf R}_0=\left(
\begin{array}{cc}
F_c(-\Omega,\theta)&-F_s(-\Omega,\theta)\\
F_s(-\Omega,\theta)&F_c(-\Omega,\theta)
\end{array}
\right),
{\bf R}_1(t')=\left(
\begin{array}{cc}
\cos(\theta-\Omega t')&-\sin(\theta-\Omega t')\\
\sin(\theta-\Omega t')&\cos(\theta-\Omega t')
\end{array}
\right).
\]
Equivalently, the above system can be rewritten as
\begin{equation}
\dot{\mbox{\boldmath$\xi$}}={\bf I}_N\otimes ({\bf J}_0-K\mu{\bf R}_0)\mbox{\boldmath$\xi$}+K\int_{0}^{\infty}g(t')[{\bf G}\otimes {\bf R}_1(t')]\mbox{\boldmath$\xi$}(t-t')dt',
\end{equation}
with the $2N$-dimensional vector $\mbox{\boldmath$\xi$}=(\xi_1,\ldots,\xi_N)^{T}$. Diagonalizing the matrix ${\bf G}$ results in a block-diagonalized variational equation
\begin{equation}
\dot{\zeta}_k(t)=\left({\bf J}_0-K\mu{\bf R}_0\right)\zeta_k(t)+K\nu_k \int_{0}^{\infty}g(t'){\bf R}_1(t')\zeta_k(t-t')dt'.
\end{equation}
Since in these equations, none of the matrices depends explicitly on time $t$, it is possible to find Floquet exponents as the eigenvalues $\Lambda\in\mathbb{C}$ of the following characteristic equation
\begin{equation}
\det\left\{{\bf J}_0-K\mu{\bf R}_0-\Lambda {\bf I}_2+K\nu_k \int_{0}^{\infty}g(t'){\bf R}_1(t')e^{-\Lambda t'}dt'\right\}=0,
\end{equation}
where ${\bf I}_2$ is a $2\times 2$ identity matrix. More explicitly, this transcendental equation has the form
\begin{equation}\label{MSF_equation}
\begin{array}{l}
\displaystyle{\Lambda^2+2\left[r_0^2+K(\mu F_c(-\Omega,\theta)-\nu_k F_c^L(-\Omega,\theta,\Lambda))\right]\Lambda+2r_0^2 K[\mu F_c(-\Omega,\theta)-
\nu_k F_c^L(-\Omega,\theta,\Lambda)]}\\\\
\displaystyle{+2 \gamma r_0^2 K[\mu F_s(-\Omega,\theta)-\nu_k F_s^L(-\Omega,\theta,\Lambda)]+K^2[\mu F_c(-\Omega,\theta)-\nu_k F_c^L(-\Omega,\theta,\Lambda)]^2}\\\\
+K^2[\mu F_s(-\Omega,\theta)-\nu_k F_s^L(-\Omega,\theta,\Lambda)]^2=0,
\end{array}
\end{equation}
where by analogy with (\ref{aux_quant}) we have introduced the quantities
\[
F_c^L(a,b,z)=\int_{0}^{\infty}g(t')\cos(at'+b)e^{-zt'}dt',\hspace{0.5cm}F_s^L(a,b,z)=\int_{0}^{\infty}g(t')\sin(at'+b)e^{-zt'}dt',
\]
where the superscript $L$ refers to these integrals representing the Laplace transform of modified kernels $g(s)\cos(as+b)$ and $g(s)\sin(as+b)$. Comparing these expressions to (\ref{aux_quant}) yields the following relations:
\[
F_c(a,b)=F_c^L(a,b,0),\hspace{0.5cm}F_s(a,b)=F_s^L(a,b,0).
\]
The equation (\ref{MSF_equation}) generalises an earlier result of Choe {\it et al.} \cite{CHO09,CHO11}, which considered the master stability function for the case of a single discrete time delay.

\section{Computation of master stability function}

\begin{figure}
\hspace{-1cm}
\epsfig{file=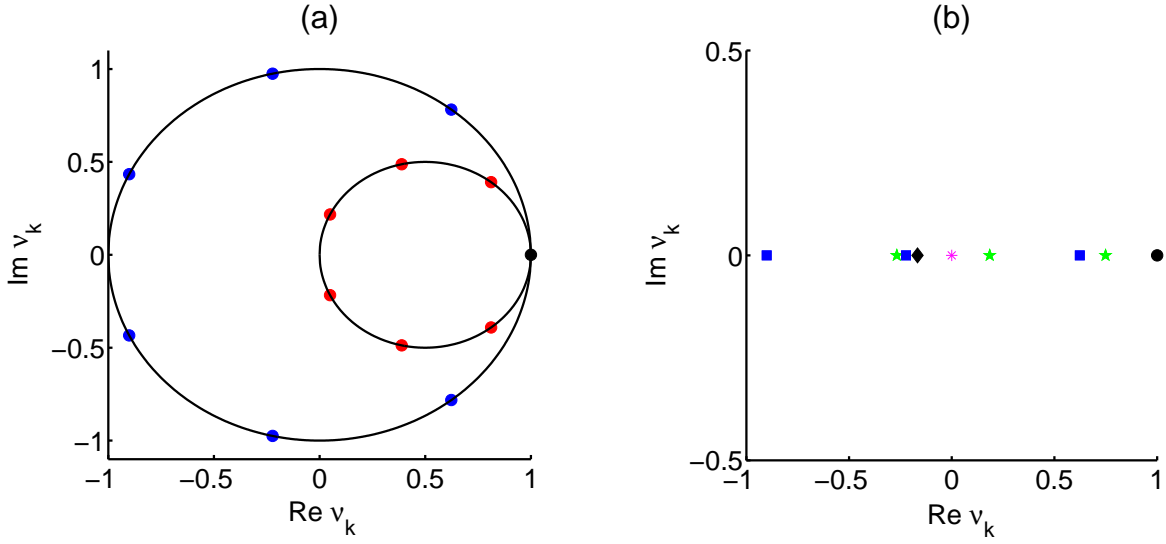,width=16.5cm}
\caption{Eigenvalues of the coupling matrices for $N=7$. (a) Uni-directional coupling without (big circle) and with (small circle) self-coupling. (b) Bi-directional coupling without (blue squares) and with (green stars) self-coupling, all-to-all coupling without (black diamond) and with (magenta asterisk) self-coupling. The longitudinal eigenvalue $\nu_0=1$ is marked by a full black circle. }\label{eigs_dia}
\end{figure}

In order to illustrate how the master stability function can be used for the analysis of stability of a synchronization manifold, we first have to specify a particular topology of the network ${\bf G}$, as well as the distributed delay kernel $g(u)$. We will consider a number of different coupling topologies, each of which is characterized by a particular spectrum of eigenvalues $\{\nu_k\}$ as shown in Fig.~\ref{eigs_dia}. Assuming that the constant row sum $\mu$ is different from zero, we normalize matrices ${\bf G}$ by dividing all elements by $\mu$, which results in a row sum equal to unity. The coupling matrices ${\bf G}$ and their eigenvalues are then given by
\begin{equation}
{\bf G}^{uni}=\left(
\begin{array}{cccccc}
0&1&0&0&\cdots &0\\
0&0&1&0&\cdots &0\\
\vdots & \vdots & \vdots & \vdots & \cdots & \vdots\\
0&0&0&0&\cdots &1\\
1&0&0&0&\cdots &0\\
\end{array}
\right),\hspace{1cm}\nu_k=e^{2\pi i k/N},\hspace{0.1cm}k=0,\ldots,N-1,
\end{equation}
for uni-directional ring coupling,
\begin{equation}
{\bf G}^{bi}=\frac{1}{2}\left(
\begin{array}{cccccc}
0&1&0&0&\cdots &1\\
1&0&1&0&\cdots &0\\
0&1&0&1&\cdots &0\\
\vdots & \vdots & \vdots & \vdots & \cdots & \vdots\\
0&0&0&0&\cdots &1\\
1&0&0&0&\cdots &0\\
\end{array}
\right),\hspace{1cm}\nu_k=\cos\frac{2\pi k}{N},\hspace{0.1cm}k=0,\ldots,N-1,
\end{equation}
for bi-directional ring coupling, and
\begin{equation}
{\bf G}^{all}_{ij}=\left\{
\begin{array}{l}
\frac{1}{N-1},\hspace{0.3cm}i\neq j,\\
0,\hspace{0.5cm}i=j,
\end{array}
\right.
,\hspace{1cm}\nu_0=1,\hspace{0.3cm}\nu_k=-\frac{1}{N-1},\hspace{0.1cm}k=1,\ldots,N-1,
\end{equation}
for all-to-all coupling. If one allows for self-feedback, then the above matrices and eigenvalues transform to
\begin{equation}
{\bf G}^{uni}_{s}=\frac{1}{2}\left(
\begin{array}{cccccc}
1&1&0&0&\cdots &0\\
0&1&1&0&\cdots &0\\
\vdots & \vdots & \vdots & \vdots & \cdots & \vdots\\
0&0&0&0&\cdots &1\\
1&0&0&0&\cdots &1\\
\end{array}
\right),\hspace{1cm}\nu_k=\frac{1}{2}\left(1+e^{2\pi i k/N}\right),\hspace{0.1cm}k=0,\ldots,N-1,
\end{equation}
for uni-directional ring coupling with self-feedback,
\begin{equation}
{\bf G}^{bi}_{s}=\frac{1}{3}\left(
\begin{array}{cccccc}
1&1&0&0&\cdots &1\\
1&1&1&0&\cdots &0\\
0&1&1&1&\cdots &0\\
\vdots & \vdots & \vdots & \vdots & \cdots & \vdots\\
0&0&0&0&\cdots &1\\
1&0&0&0&\cdots &1\\
\end{array}
\right),\hspace{1cm}\nu_k=\frac{1}{3}\left(1+2\cos\frac{2\pi k}{N}\right),\hspace{0.1cm}k=0,\ldots,N-1,
\end{equation}
for bi-directional ring coupling with self-feedback, and
\begin{equation}
{\bf G}^{all}_{s,ij}=\frac{1}{N},\hspace{0.3cm}i,j=1,\ldots, N;\hspace{1cm}\nu_0=1,\hspace{0.3cm}\nu_k=0,\hspace{0.1cm}k=1,\ldots,N-1,
\end{equation}
for all-to-all coupling with self-feedback.

All of the above-considered coupling matrices of uni-directional, bi-directional and all-to-all couplings with or without self-feedback are special cases of an important class of networks, whose topology is described by the so-called {\it circulant matrices}, which are $N\times N$ matrices of the form
\begin{equation}
{\bf C}=\left(
\begin{array}{ccccc}
c_1& c_N &\cdots &c_3&c_2\\
c_2 & c_1 & c_N & \cdots & c_3\\
\vdots & c_2 & c_1 & \ddots & \vdots\\
c_{N-1} & &\ddots & \ddots &c_N\\
c_N & c_{N-1} & \cdots & c_2 & c_1
\end{array}
\right).
\end{equation}
The eigenvalues of this general matrix ${\bf C}$ are given by
\begin{equation}
\nu_k=c_1+c_N \omega_k+c_{N-1}\omega_k^2+\ldots+c_2\omega_k^{N-1}=\sum_{m=1}^{N}c_m\omega_k^{(1-m)+N\mbox{ }{\rm mod }\mbox{ }N},\hspace{0.5cm}k=0,\ldots,N-1,
\end{equation}
where $\omega_k=\exp(2\pi ik/N)$.

In terms of the distributed delay kernel, we consider two practically important choices of the delay kernel: the uniformly distributed kernel and the gamma distributed delay kernel. The uniformly distributed delay kernel is given by
\begin{equation}\label{UKer}
g(u)=\left\{
\begin{array}{l}
\displaystyle{\frac{1}{2\rho} \hspace{1cm}\mbox{for }\tau-\rho\leq u\leq \tau+\rho,}\\\\
0\hspace{1cm}\mbox{elsewhere.}
\end{array}
\right.
\end{equation}
This distribution has the mean time delay 
\[
\tau_{m}\equiv<\tau>=\int_{0}^{\infty}ug(u)du=\tau,
\]
and the variance
\begin{equation}\label{VD}
\displaystyle{Var=\int_{0}^{\infty}(u-\tau_m)^{2}g(u)du=\frac{\rho^{2}}{3}.}
\end{equation}
The uniformly distributed delay kernel (\ref{UKer}) has been successfully used in a number of different contexts, including models of traffic dynamics with delayed driver response \cite{SAN07}, stem cell dynamics \cite{Adimy}, time-delayed feedback control \cite{GU1}, and genetic regulation \cite{Barrio}.

The second example we consider is that of the gamma distribution, which can be written as
\begin{equation}\label{GDdef}
g(u)=\frac{u^{p-1}\alpha^{p}e^{-\alpha u}}{\Gamma(p)},
\end{equation}
with $\alpha,p\geq 0$, and $\Gamma(p)$ being the Euler gamma function defined by
\[
\Gamma(p)=\int_0^{\infty}x^{p-1}e^{-x}dx,
\]
and satisfying $\Gamma(0)=1$ and $\Gamma(p+1)=p\Gamma(p)$ for any real $p$. For integer powers $p$, the delay distribution (\ref{GDdef}) can be equivalently written as
\begin{equation}\label{GD}
g(u)=\frac{u^{p-1}\alpha^{p}e^{-\alpha u}}{(p-1)!}.
\end{equation}
For $p=1$ this is simply an exponential distribution (also called a {\it weak delay kernel}) with the maximum contribution to the coupling coming from the instantaneous values of the variables $z_{k}$.

For $p>1$ (known as {\it strong delay kernel} in the case $p=2$), the biggest influence on the coupling at any moment of time $t$ is from the values of $z_{k}$ at $t-(p-1)/\alpha$. The delay distribution (\ref{GD}) has the mean time delay
\begin{equation}\label{taum}
\displaystyle{\tau_{m}=\int_{0}^{\infty}ug(u)du=\frac{p}{\alpha},}
\end{equation}
and the variance
\[
\displaystyle{Var=\int_{0}^{\infty}(u-\tau_m)^{2}g(u)du=\frac{p}{\alpha^2}}.
\]
The gamma distributed delay kernel (\ref{GD}) was originally analysed in models of population dynamics \cite{Blythe,Cush,CZ}, and has subsequently been used to study machine tool vibrations \cite{STE98a}, intracellular dynamics of HIV infection \cite{Mit}, traffic dynamics with delayed driver response \cite{SNA07}, and control of objects over wireless communication networks \cite{Roe}.

Once the distributed delay kernel is fixed, one can explicitly compute the functions $F_{c,s}$ and $F_{c,s}^L$ as follows. For the uniform distribution (\ref{UKer}), they are given by
\[
F_c(-\Omega,\theta)=\frac{1}{\rho\Omega}\sin(\rho\Omega)\cos(\theta-\Omega\tau),\hspace{0.5cm}F_s(-\Omega,\theta)=\frac{1}{\rho\Omega}\sin(\rho\Omega)\sin(\theta-\Omega\tau),
\]
and the case of a discrete time delay can be recovered by setting $\rho=0$. For the weak distribution kernel (\ref{GD}) with $p=1$, we have
\[
F_c(-\Omega,\theta)=\alpha\frac{\alpha\cos\theta+\Omega\sin\theta}{\alpha^2+\Omega^2},\hspace{0.5cm}F_s(-\Omega,\theta)=\alpha\frac{\alpha\sin\theta-\Omega\cos\theta}{\alpha^2+\Omega^2},
\]
and similarly, for the strong delay kernel (\ref{GD}) with $p=2$:
\[
F_c(-\Omega,\theta)=\alpha^2\frac{(\alpha^2-\Omega^2)\cos \theta+2\alpha\Omega\sin\theta}{(\alpha^2+\Omega^2)^2},\hspace{0.5cm}F_s(-\Omega,\theta)=\alpha^2\frac{(\alpha^2-\Omega^2)\sin\theta-2\alpha\Omega\cos\theta}{(\alpha^2+\Omega^2)^2}.
\]
In the same way, Laplace transforms with the modified kernels give
\[
\begin{array}{l}
\displaystyle{F_c^L(-\Omega,\theta,\Lambda)=\frac{e^{-\Lambda\tau}}{2\rho(\Lambda^2+\Omega^2)}\Big[\Lambda e^{\Lambda\rho}\cos[\theta-\Omega(\tau-\rho)]-
\Lambda e^{-\Lambda\rho}\cos[\theta-\Omega(\tau+\rho)]}\\\\
\displaystyle{+\Omega e^{\Lambda\rho}\sin[\theta-\Omega(\tau-\rho)]-\Omega e^{-\Lambda\rho}\sin[\theta-\Omega(\tau+\rho)]\Big],}\\\\
\displaystyle{F_s^L(-\Omega,\theta,\Lambda)=\frac{e^{-\Lambda\tau}}{2\rho(\Lambda^2+\Omega^2)}\Big[\Lambda e^{\Lambda\rho}\sin[\theta-\Omega(\tau-\rho)]-
\Lambda e^{-\Lambda\rho}\sin[\theta-\Omega(\tau+\rho)]}\\\\
\displaystyle{+\Omega e^{-\Lambda\rho}\cos[\theta-\Omega(\tau+\rho)]-\Omega e^{\Lambda\rho}\cos[\theta-\Omega(\tau-\rho)]\Big],}
\end{array}
\]
for the uniform distribution kernel,
\[
\displaystyle{F_c^L(-\Omega,\theta,\Lambda)=\alpha\frac{(\alpha+\Lambda)\cos\theta+\Omega\sin\theta}{(\alpha+\Lambda)^2+\Omega^2},\hspace{0.5cm}F_s^L(-\Omega,\theta,\Lambda)=\alpha\frac{(\alpha+\Lambda)\sin\theta-\Omega\cos\theta}{(\alpha+\Lambda)^2+\Omega^2},}
\]
for the weak distribution kernel, and
\[
\begin{array}{l}
\displaystyle{F_c^L(-\Omega,\theta,\Lambda)=\alpha^2\frac{\left[(\alpha+\Lambda)^2-\Omega^2\right]\cos \theta+2(\alpha+\Lambda)\Omega\sin\theta}{\left[(\alpha+\Lambda)^2+\Omega^2\right]^2},}\\\\
\displaystyle{F_s(-\Omega,\theta)=\alpha^2\frac{\left[(\alpha+\Lambda)^2-\Omega^2\right]\sin\theta-2(\alpha+\Lambda)\Omega\cos\theta}{\left[(\alpha+\Lambda)^2+\Omega^2\right]^2},}
\end{array}
\]
for the strong distribution kernel.

\begin{figure}
\epsfig{file=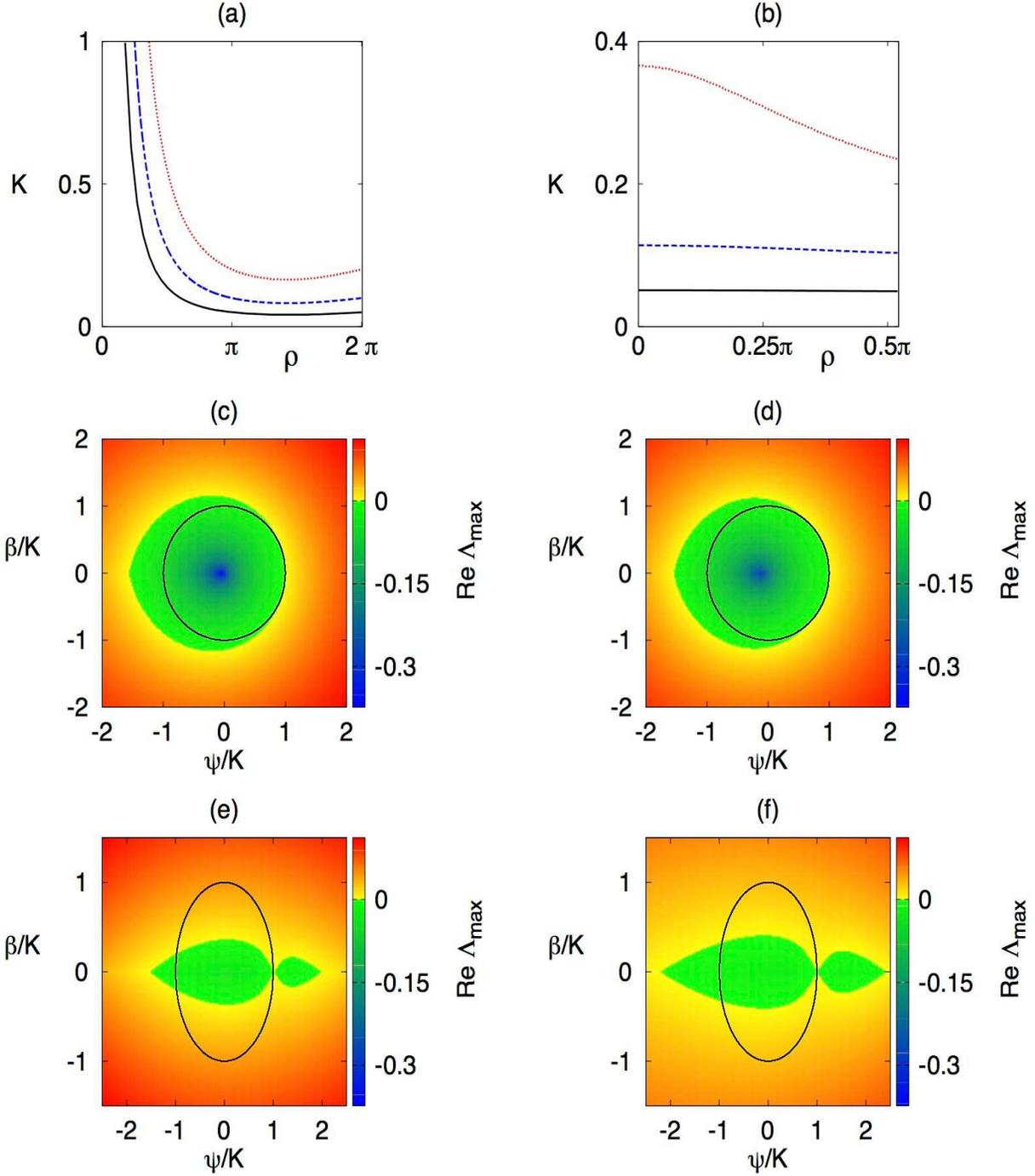,width=17cm,height=19cm}
\caption{(a),(b) Existence of synchronized solution and (c)-(f) master stability function Re$[\Lambda_{\rm max}(\psi,\beta)]$ for coupled Stuart-Landau oscillators with uniform delay coupling. Parameter values are $\omega=1$, $\gamma=0$, $\theta=0$, $\mu=1$. (a) $\tau=2\pi$, (b) $\tau=0.52\pi$, $\lambda=0.05$ (black solid), $\lambda=0.1$ (blue dashed), $\lambda=0.2$ (red dotted), fully synchronized state exists below the corresponding curves. (c)-(d) Master stability function, $\lambda=0.1$, $K=0.3$, $\tau=2\pi$, (c) $\rho=0$, (d) $\rho=1.49$. (e)-(f) Master stability function, $\lambda=0.1$, $K=0.08$, $\tau=0.52\pi$, (e) $\rho=0$, (f) $\rho=0.5\pi$. All eigenvalues of the coupling matrix lie on or inside the black circle.}\label{msf_uni}
\end{figure}

As a first step in the analysis of stability of the fully synchronized solution (\ref{r0eq}), one should note that this state does not always exist for arbitrary values of the coupling strength and phase and parameters of the delay distribution. The regions of parameter space where this state exists are determined by the condition $r_0^2=\lambda+K\mu\left[F_c(-\Omega,\theta)-\cos\theta\right]\geq 0$. Figures~\ref{msf_uni} (a) and (b) illustrate such regions for a uniform delay distribution with different choices of the mean time delay, and suggest a natural conclusion that increasing the linear rate $\lambda$ increases the range of values of the coupling strength $K$ for which a fully synchronized solution exists. Once the parameter regions where the fully synchronized state exists have been identified, we choose particular values of the coupling strength $K$ and the mean time delay $\tau$, and then compute the master stability function Re$[\Lambda_{\rm max}(\psi,\beta)]$ by solving the equation (\ref{MSF_equation}), as shown in Figures~\ref{msf_uni} (c) to (f). Stability for individual coupling topologies is verified by checking that Re$[\Lambda_{\rm max}(\psi,\beta)]<0$ for all transverse eigenvalues $\nu_k$ of the coupling matrix ${\bf G}$ with $K\nu_k=\psi+i\beta$. These Figures suggest that for the same value of the mean time delay, a uniform distribution with a larger width of the distribution results in the same or a slightly larger region of stable synchronization in the $(\psi,\beta)$ plane. Note, however, that this region might also shrink, according to the choice of the mean delay. In the particular case of mean time delay $\tau=2\pi$, all coupling topologies result in a stable synchronization manifold regardless of the width of the distribution $\rho$. On the other hand, for $\tau=0.52\pi$, the synchronization manifold is always unstable for uni-directional coupling (both with and without self-feedback) and is always stable for bi-directional and all-to-all couplings without and with self-feedback. The values of the coupling strength $K$ were taken to be the same as in the earlier work on synchronization of coupled Stuart-Landau oscillators \cite{CHO09,CHO11} to illustrate the role played by the width of the delay distribution.

\begin{figure}
\epsfig{file=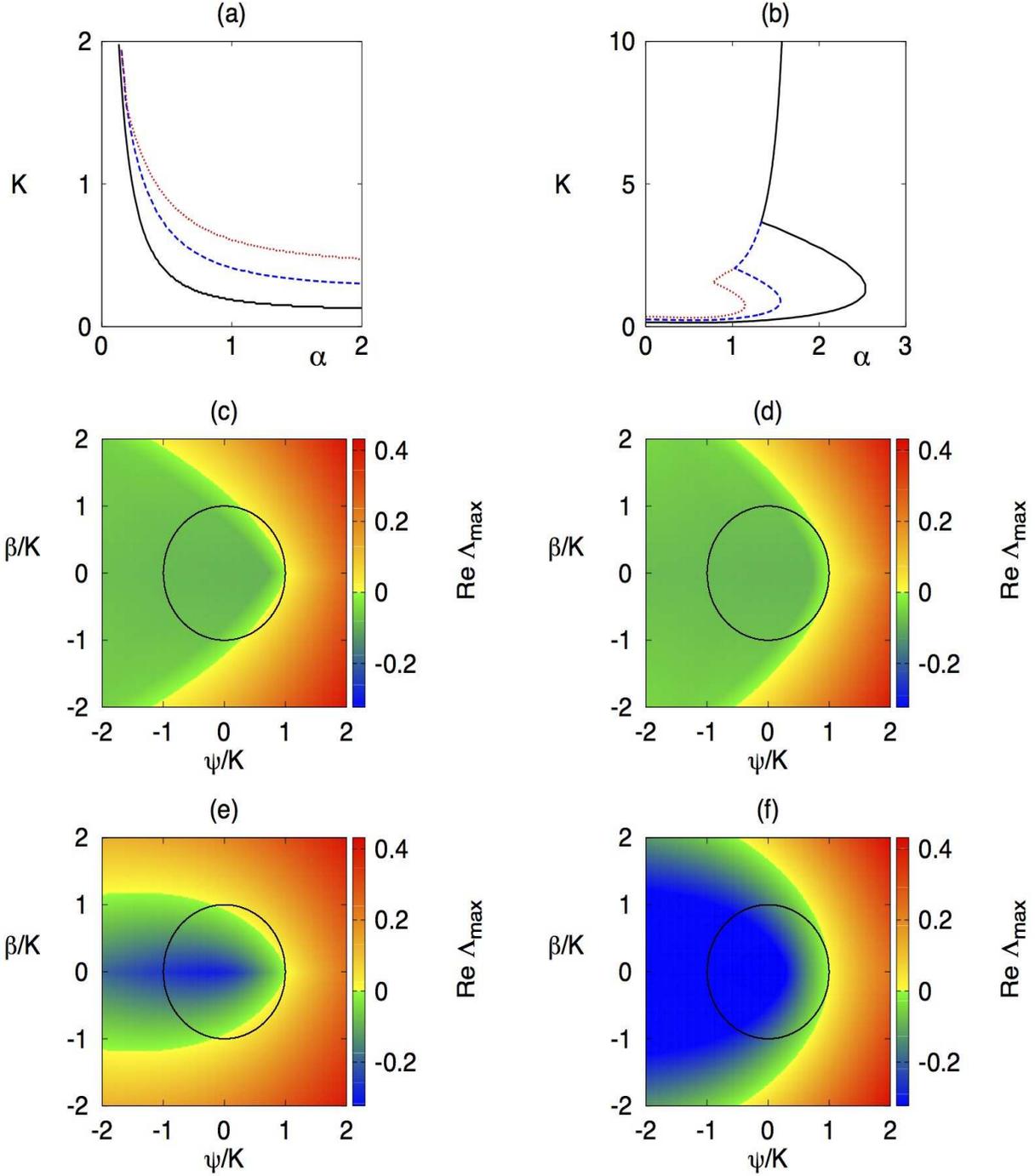,width=17cm,height=19cm}
\caption{Same as Fig.~\ref{msf_uni} with gamma distributed delay coupling. Parameter values are $\omega=1$, $\gamma=0$, $\theta=0$, $\mu=1$. (a) weak kernel $(p=1)$, $\lambda=0.1$ (black solid), $\lambda=0.25$ (blue dashed), $\lambda=0.4$ (red dotted), fully synchronized state exists below the corresponding curve; (b) strong kernel $(p=2)$, $\lambda=0.15$ (black solid), $\lambda=0.25$ (blue dashed), $\lambda=0.35$ (red dotted), fully synchronized state exists below the corresponding curve. (c)-(d) weak kernel $(p=1)$, $\lambda=0.25$, $K=0.5$, (c) $\alpha=0.8$. (d) $\alpha=1$. (e)-(f) strong kernel $(p=2)$,  (e) $\alpha=1.5$. (f) $\alpha=3$. All eigenvalues of the coupling matrix lie on or inside the black circle.}\label{msf_gamma}
\end{figure}

Figure~\ref{msf_gamma} illustrates the computation of regions of existence of the fully synchronized solution, and the master stability function for the gamma delay distribution in the case of $p=1$ and $p=2$. Similarly to the case of the uniform delay distribution, increasing the bifurcation parameter $\lambda$ leads to an increase in the region of parameters, where the fully synchronized solutions exist. For both weak ($p=1$) and strong ($p=2$) gamma distribution kernels, decreasing the average time delay (which is equivalent to increasing the parameter $\alpha$ in the gamma distribution) leads to a wider region of stability. Unlike the case of the uniform distribution kernel, varying $\alpha$ can lead to (de)stabilization of certain coupling topologies: while for small values of $\alpha$ only the bi-directional and all-to-all couplings can be stable, as $\alpha$ increases the uni-directional coupling is also stabilized.

\section{Stability of cluster and splay states}

So far, we have considered identical or zero-lag synchronization, where all oscillators have exactly the same amplitudes and phases. It is possible to use a similar approach for studying other types of synchronization, with one particularly important type being that of cluster and splay states. In this case, all oscillators still have the same amplitude $r_{0,m}$ and the same common frequency $\Omega_m$, but rather than having equal phases, now they have constant differences in their phases:
\begin{equation}\label{splay}
r_k(t)=r_{0,m},\hspace{0.5cm}\phi_k(t)=\Omega_m t+k\Delta\phi_m,\hspace{0.5cm}k=1,\ldots,N,
\end{equation}
with $\Delta\phi_m=2\pi m/N$, where $m=0,\ldots,N-1$. The case $m=0$ corresponds to identical (or in-phase) synchronization considered earlier, while $m=1,2,...,N-1$ correspond to cluster and splay states. The cluster number $N_c={\rm lcm}(m,N)/m$, where ${\rm lcm(\cdot,\cdot)}$ denotes the least common multiple, determines how many clusters of oscillators exist, with $N_c=N$ corresponding to a splay state \cite{CHO09,ZIL07}.

As it has been shown in \cite{CHO11}, the cluster solution (\ref{splay}) exists if and only if the following conditions hold
\begin{equation}\label{splay_cond}
\sum_{j=1}^N G_{kj}\cos\left[(j-k)\Delta\phi_m\right]=G^c_m={\rm const},\hspace{0.5cm}
\sum_{j=1}^N G_{kj}\sin\left[(j-k)\Delta\phi_m\right]=G^s_m={\rm const},
\end{equation}
independently of $k$. Although these conditions can be satisfied by different kinds of matrices (see \cite{CHO11} for a detailed discussion), importantly, they are satisfied by all circulant matrices, of which many standard network topologies are special cases. Provided that the above conditions hold, the common radius $r_{0,m}$ and the common frequency $\Omega_m$ are determined by the following equations
\begin{equation}
\begin{array}{l}
r_{0,m}^2=\lambda+K\mu[\mathcal{G}^c_m(-\Omega_m,\theta)-\cos\theta],\\\\
\Omega_m=\omega-\gamma r_{0,m}^2+K\mu[\mathcal{G}^s_m(-\Omega_m,\theta)-\sin\theta].
\end{array}
\end{equation}
where
\[
\begin{array}{l}
\mathcal{G}^c_m(-\Omega_m,\theta)=G^c_m F_c(-\Omega_m,\theta)-G^s_m F_s(-\Omega_m,\theta),\\\\
\mathcal{G}^s_m(-\Omega_m,\theta)=G^s_m F_c(-\Omega_m,\theta)+G^c_m F_s(-\Omega_m,\theta).
\end{array}
\]

Using the ansatz $r_k(t)=r_{0,m}[1+\delta r_k(t)]$, $\phi_k(t)=\Omega_m t+k\Delta\phi_m+\delta\phi_k(t)$, the variational equation for linearization near the cluster state (\ref{splay}) can be found as follows
\begin{equation}
\dot{\xi}_k=({\bf J}_{0,m}-K\mu {\bf \mathcal{G}})\xi_k+ K\sum_{j=1}^{N}G_{kj}\int_0^{\infty}g(t'){\bf R}_{j,m}(t')\xi_{j}(t-t')dt',\hspace{0.5cm}k=1,\ldots,N,
\end{equation}
where $\xi_k=(\delta r_k,\delta \phi_k)^{T}$, and we have also introduced the matrices
\[
{\bf J}_{0,m}=\left(
\begin{array}{cc}
-2r_{0,m}^2&0\\
-2\gamma r_{0,m}^2&0
\end{array}
\right),\hspace{0.2cm}
{\bf \mathcal{G}}=\left(
\begin{array}{cc}
\mathcal{G}^c_m(-\Omega_m,\theta) &-\mathcal{G}^s_m(-\Omega_m,\theta)\\
\mathcal{G}^s_m(-\Omega_m,\theta) & \mathcal{G}^c_m(-\Omega_m,\theta)
\end{array}
\right),\hspace{0.2cm}
{\bf R}_{j,m}(t')=\left(
\begin{array}{cc}
\cos\Phi_j^m&-\sin\Phi_j^m\\
\sin\Phi_j^m&\cos\Phi_j^m
\end{array}
\right),
\]
and the phase difference $\Phi_j^m=-\Omega_m t'+(j-k)\Delta\phi_m+\theta$. Using the same approach as before, one can recast this equation in the form
\begin{equation}
\dot{\mbox{\boldmath$\xi$}}={\bf I}_N\otimes ({\bf J}_{0,m}-K\mu {\bf \mathcal{G}})\mbox{\boldmath$\xi$}+K\int_0^{\infty}g(t')[{\bf G}\otimes {\bf R}_{j,m}]\mbox{\boldmath$\xi$}(t-t')dt',
\end{equation}
with $\mbox{\boldmath$\xi$}=(\xi_1,\ldots,\xi_N)^{T}$. To make further progress in the analysis of stability of the cluster state, one has to diagonalize the matrix ${\bf G}$ and study Floquet exponents corresponding to transverse eigenvalues of this matrix. However, unlike the case of identical synchronization $m=0$, this is in general impossible, unless the matrix ${\bf G}$ has certain properties which make the matrix ${\bf R}_{j,m}$ independent of $k$. This is the case for all topologies considered in the previous section, which are represented by circulant matrices.

To illustrate the computation of the master stability function for cluster states, we consider the network of the Stuart-Landau oscillators with uni-directional ring coupling: $G_{k,k+1}=G_{N,1}=1=\mu$, and all other $G_{kj}=0$. In this case, we have $\Phi_j^m=-\Omega_m t'+\Delta\phi_m+\theta$, and the matrix ${\bf R}_{j,m}$ reduces to
\[
{\bf R}_{j,m}(t')={\bf R}_1(t')=\left(
\begin{array}{cc}
\cos(\theta-\Omega_m t'+\Delta\phi_m)&-\sin(\theta-\Omega_m t'+\Delta\phi_m)\\
\sin(\theta-\Omega_m t'+\Delta\phi_m)&\cos(\theta-\Omega_m t'+\Delta\phi_m)
\end{array}
\right).
\]
Since this matrix does not depend on $k$, one can use the same approach as in the case of a fully synchronized solution to diagonalize the matrix ${\bf G}$, which results in the block-diagonalized variational equation
\[
\dot{\zeta}_k(t)=({\bf J}_{0,m}-K\mu {\bf \mathcal{G}})\zeta_k(t)+K\nu_k\int_{0}^{\infty}g(t'){\bf R}_1(t')\zeta(t-t')dt',
\]
where the eigenvalues $\nu_k$ of the matrix ${\bf G}$ are explicitly given by $\displaystyle{\nu_k=e^{2ik\pi/N}}$, $k=0,1,\ldots,N-1$. Since all terms in the above equation are independent of time, the Floquet exponents can be found as the eigenvalues $\Lambda$ of the characteristic equation
\begin{equation}
\det\left\{{\bf J}_{0,m}-K\mu{\bf \mathcal{G}}-\Lambda {\bf I}_2+K\nu_k \int_{0}^{\infty}g(t'){\bf R}_1(t')e^{-\Lambda t'}dt'\right\}=0,
\end{equation}
where ${\bf I}_2$ is a $2\times 2$ identity matrix. More explicitly, this transcendental equation has the form
\begin{equation}
\begin{array}{l}
\displaystyle{\Lambda^2+2\left[r_{0,m}^2+K(\mu \mathcal{G}^c_m(-\Omega_m,\theta)-\nu_k F_c^L(-\Omega_m,\theta+\Delta\phi_m,\Lambda))\right]\Lambda}\\\\
\displaystyle{+2r_{0,m}^2 K\Big[\mu \mathcal{G}^c_m(-\Omega_m,\theta)-\nu_k F_c^L(-\Omega_m,\theta+\Delta\phi_m,\Lambda)+
\gamma(\mu \mathcal{G}^s_m(-\Omega_m,\theta)-\nu_k F_s^L(-\Omega_m,\theta+\Delta\phi_m,\Lambda))\Big]}\\\\
\displaystyle{+K^2[\mu \mathcal{G}^c_m(-\Omega_m,\theta)-\nu_k F_c^L(-\Omega_m,\theta+\Delta\phi_m,\Lambda)]^2
+K^2[\mu \mathcal{G}^s_m(-\Omega_m,\theta)-\nu_k F_s^L(-\Omega_m,\theta+\Delta\phi_m,\Lambda)]^2}\\\\
\displaystyle{+K^2[\mu \mathcal{G}^s_m(-\Omega_m,\theta)-\nu_k F_s^L(-\Omega_m,\theta+\Delta\phi_m,\Lambda)]^2=0.}
\end{array}
\end{equation}
In the case of a single discrete time delay $g(u)=\delta(u-\tau)$, this equation reduces to a case investigated by Choe {\it et al.} \cite{CHO09,CHO11}.

\begin{figure}
\epsfig{file=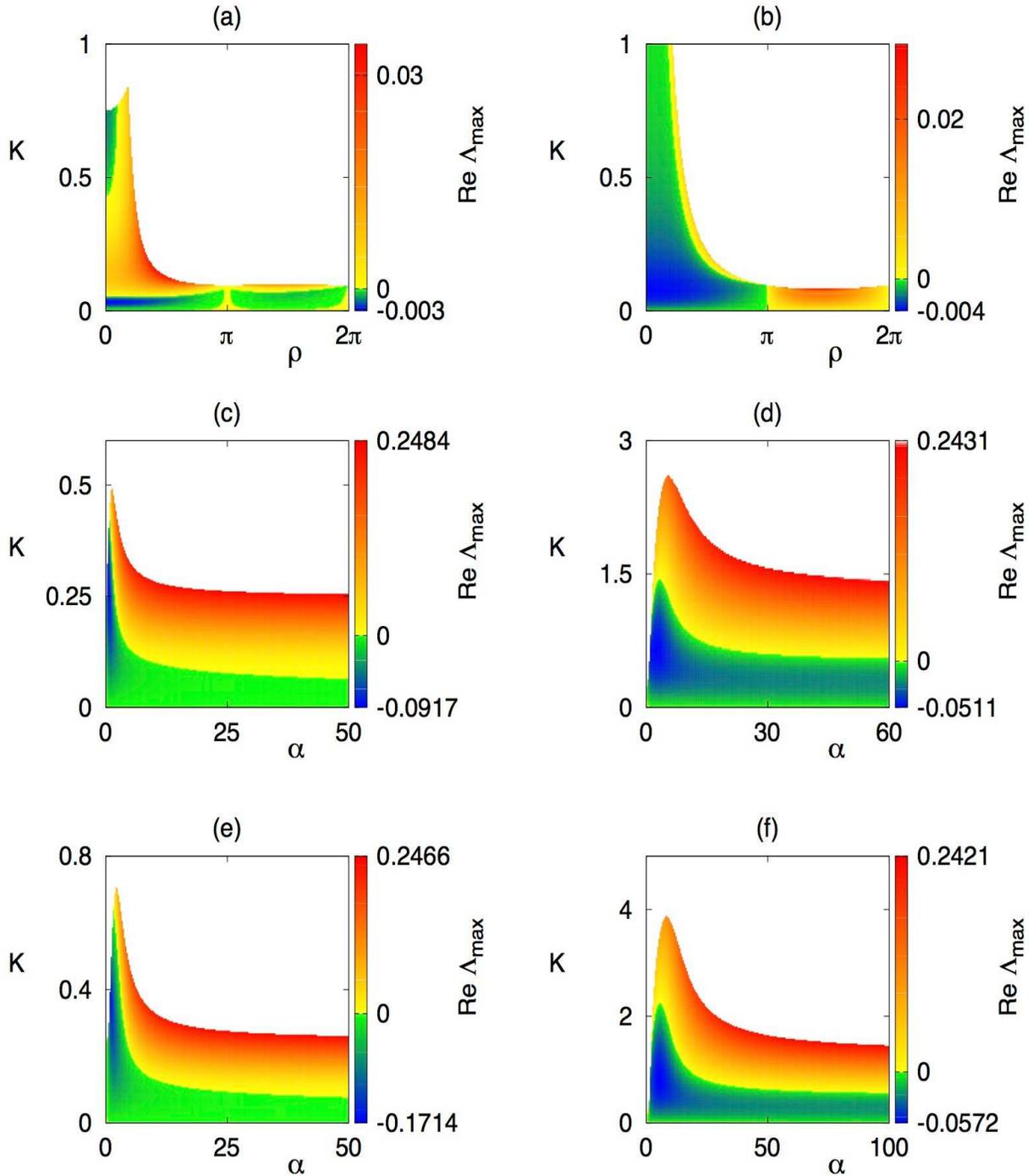,width=17cm,height=19cm}
\caption{Master stability function for a splay state with $m=1$ for a ring of uni-directionally coupled Stuart-Landau oscillators. Parameter values are $\omega=1$, $\gamma=0$, $\theta=0$, $\mu=1$. (a)-(b) uniform delay distribution, $\tau=2\pi$, $\lambda=0.1$, (a) $N=4$,  (b) $N=10$. (c)-(d) weak kernel $(p=1)$, $\lambda=0.25$, (c) $N=4$, (d) $N=10$. (e)-(f) strong kernel $(p=2)$, $\lambda=0.25$, (e) $N=4$, (f) $N=10$. The splay state does not exist in the white region.}\label{splay_plot}
\end{figure}

Figure~\ref{splay_plot} illustrates the computation of the master stability function of the cluster state for uni-directional ring coupling and uniform or gamma delay distributions. For both of these delay distributions, an increase in the number of oscillators appears to increase the range of coupling strengths for which the cluster state exists, and also proportionally increase the region of stability of this state. In the case of the gamma distributed delay kernel, one can note that whenever it exists, the cluster state is stable for lower values of the coupling strength across the full range of admissible values of parameter $\alpha$.

\section{Discussion}

In this paper we have shown how the framework of the master stability function for the computation of stability of zero-lag and cluster synchronized states in systems of coupled oscillators can be generalized to the case of distributed delay coupling. To illustrate how the master stability function framework can be used for studying the stability of synchronized states, we have considered the example of a network of coupled Stuart-Landau oscillators, which is a normal form of a supercritical Hopf bifurcation. In this example, computation of the master stability function, i.e., the Floquet exponents, reduces to finding the corresponding eigenvalues of a characteristic equation, which can be done semi-analytically in terms of system parameters, topological eigenvalues of the coupling matrix, and the coupling parameters, where the Laplace transform of the delay kernel enters.  Calculations for the case of uniform and gamma distributed delay kernels show that increasing the width of the delay distribution (for a uniform kernel) or reducing the mean time delay (for a gamma distributed delay kernel) leads often to a larger region of stability, thus allowing the fully synchronized state to be stable for a larger class of coupling topologies.

Although the analysis presented in this paper has focused upon the case of linear coupling between different oscillators in the network, it is straightforward to generalize our results on the stability of zero-lag and cluster synchronization to systems with nonlinear coupling \cite{PHT05}, since the master stability equation is still a linear variational equation involving the appropriate delay distribution kernel. Recent studies of amplitude death in systems of delay-coupled oscillators \cite{KBS11,KBS13,ZOU11} have highlighted the important role played by the coupling phase in determining the regions of possible suppression of oscillations. Due to substantial analytical headway provided by considering Stuart-Landau oscillators, it would be insightful to use the master stability framework developed in this paper to investigate the effects of the coupling phase on existence and stability of synchronized solutions in networks of such oscillators.

\section*{Acknowledgements}

This work was partially supported by DFG in the framework of SFB 910: {\em Control of self-organizing nonlinear
systems: Theoretical methods and concepts of application}. YK and KB gratefully acknowledge the hospitality of the Institut f\"ur Theoretische Physik, TU Berlin, where part of this work was completed. ES is grateful to Carolin Wille for helpful discussions. The authors thank the anonymous referees for helpful comments and suggestions that have helped to improve the presentation.


\begin{thebibliography}{70}

\bibitem{AB} R. Albert and A.-L. Barab\'asi, Rev. Mod. Phys. {\bf 74}, 47 (2002).

\bibitem{BB} A.-L. Barab\'asi and E. Bonabeau, Sci. Am. {\bf 288}, 60 (2003).

\bibitem{New} M.E.J. Newman, SIAM Review {\bf 45}, 167 (2003).

\bibitem{WS} D.J. Watts and S.H. Strogatz, Nature {\bf 393}, 440 (1998).

\bibitem{BOC06} S. Boccaletti, V. Latora, Y. Moreno, M. Chavez, and D.U. Hwang, Phys. Rep. {\bf 424}, 175 (2006).

\bibitem{PRK01} A. Pikovsky, M. Rosenblum, and J. Kurths, {\it Synchronization: a universal concept in nonlinear sciences} (CUP, Cambridge, 2001).

\bibitem{PC98} L.M. Pecora and T.L. Carroll, Phys. Rev. Lett. {\bf 80}, 2109 (1998).

\bibitem{ES} M.G. Earl and S.H. Strogatz, Phys. Rev. E {\bf 67}, 036204 (2003).

\bibitem{DVEDF08} O. D'Huys, R. Vicente, T. Erneux, J. Danckaert, and I. Fischer, Chaos {\bf 18}, 037116 (2008).

\bibitem{K09} W. Kinzel, A. Englert, G. Reents, M. Zigzag, and I. Kanter, Phys. Rev. E {\bf 79}, 056207 (2009).

\bibitem{DVDF10} O. D'Huys, R. Vicente, J. Danckaert, and I. Fischer, Chaos {\bf 20}, 043127 (2010).

\bibitem{EKA10} A. Englert, W. Kinzel, Y. Aviad, M. Butkovski, I. Reidler, M. Zigzag, I. Kanter, and M. Rosenbluh, Phys. Rev. Lett. {\bf 104}, 114102 (2010).

\bibitem{FYDS10}  V. Flunkert, S. Yanchuk, T. Dahms, and E. Sch\"oll, Phys. Rev. Lett. {\bf 105}, 254101 (2010).

\bibitem{ATA10} F.M. Atay (Ed.), {\it Complex Time-Delay Systems} (Springer, New York, 2010).

\bibitem{EHKK11} A. Englert, S. Heiligenthal, W. Kinzel, and I. Kanter, Phys. Rev. E {\bf 83}, 046222 (2011).

\bibitem{HHFS11} K. Hicke, O. D'Huys, V. Flunkert, E. Sch\"oll, I. Danckaert, and I. Fischer, Phys. Rev. E {\bf 83}, 056211 (2011).

\bibitem{LDHS} J. Lehnert, T. Dahms, P. H\"ovel, and E. Sch\"oll, Europhys. Lett. {\bf 96}, 60013 (2011).

\bibitem{FLU13} V. Flunkert, I. Fischer, and E. Sch{\"o}ll (Eds.), {\it Dynamics, control and information in delay-coupled systems}. Theme Issue of Phil. Trans. R. Soc. London A {\bf 371}, 20120465 (2013).

\bibitem{SOR13} M.~C. Soriano, J. Garc{\'i}a-Ojalvo, C.~R. Mirasso, and I. Fischer, Rev.~Mod.~Phys. {\bf 85},  421 (2013).

\bibitem{ZAK13a} A. Zakharova, I. Schneider, Y.~N. Kyrychko, K.~B. Blyuss, A. Koseska, B. Fiedler, and E. Sch{\"o}ll, Europhys.~Lett. {\bf 104}, 50004  (2013).

\bibitem{CHO09} C.-U. Choe, T. Dahms, P. H\"ovel, and E. Sch\"oll, Phys. Rev. E {\bf 81}, 025205(R) (2010).

\bibitem{CHO11} C.-U. Choe, T. Dahms, P. H\"ovel, and E. Sch\"oll, in {\it Proceedings of the
Eighth AIMS International Conference on Dynamical Systems, Differential
Equations and Applications} (American Institute of Mathematical Sciences,
Springfield, 2011), pp. 292-301.

\bibitem{CHO13} C.-U. Choe, H. Jang, V. Flunkert, T. Dahms, P. H\"ovel, and E. Sch\"oll, Dyn. Syst. {\bf 28}, 15 (2013).

\bibitem{DAH12} T. Dahms, J. Lehnert, and E. Sch{\"o}ll, Phys. Rev.~E {\bf 86},  016202 (2012).

\bibitem{DJD04} M. Dhamala, V.K. Jirsa, and M. Ding, Phys. Rev. Lett. {\bf 92}, 074104 (2004).

\bibitem{WIL13} C.~R.~S. Williams, T.~E. Murphy, R. Roy, F. Sorrentino, T. Dahms, and E. Sch{\"o}ll, Phys. Rev. Lett. {\bf 110},  064104 (2013).

\bibitem{Hunt1} D. Hunt, G. Korniss, and B. K. Szymanski, Phys. Rev. Lett. {\bf 105}, 068701 (2010). 

\bibitem{Hunt2} D. Hunt, B. K. Szymanski, and G. Korniss, Phys. Rev. E {\bf 86}, 056114 (2012). 

\bibitem{GU1} A. Gjurchinovski and V. Urumov, Europhys. Lett. {\bf 84}, 40013 (2008).

\bibitem{GU2} A. Gjurchinovski and V. Urumov, Phys. Rev. E {\bf 81}, 016209 (2010).

\bibitem{GJU14} A. Gjurchinovski, A. Zakharova, and E. Sch{\"o}ll, Phys. Rev. E {\bf 89}, 032915  (2014).

\bibitem{ATA03} F. Atay, Phys. Rev. Lett. {\bf 91}, 094101 (2003).

\bibitem{CJ09} S.A. Campbell and R. Jessop, Math. Model. Nat. Phenom. {\bf 4}, 1 (2009).

\bibitem{CHO14} C.~U. Choe, R.-S. Kim, H. Jang, P. H{\"o}vel, and E. Sch{\"o}ll, Int. J. Dynam. Control {\bf 2},  2 (2014).

\bibitem{KK11} G. Kiss and B. Krauskopf, Dyn. Syst. {\bf 26}, 85 (2011).

\bibitem{GJU13} A. Gjurchinovski, T. J{\"u}ngling, V. Urumov, and E. Sch{\"o}ll, Phys. Rev. E {\bf 88},  032912  (2013).

\bibitem{MVAN05} W. Michiels, V. Van Assche, and S.-I. Niculescu, IEEE Trans. Automat. Contr. {\bf 50}, 493 (2005).

\bibitem{BK10} K.B. Blyuss and Y.N. Kyrychko, Bull. Math. Biol. {\bf 72}, 490 (2010). 

\bibitem{FT10} T. Faria and S. Trofimchuk, Nonlinearity {\bf 23}, 2457 (2010).

\bibitem{GS03} S.A. Gourley and J.W.-H. So, Proc. R. Soc. Edinburgh {\bf 133}, 527 (2003).

\bibitem{SAN07} R. Sipahi, F.M. Atay, and S.-I. Niculescu, SIAM J. Appl. Math. {\bf 68}, 738 (2007).

\bibitem{ETF05} C.W. Eurich, A. Thiel, and L. Fahse, Phys. Rev. Lett. {\bf 94}, 158104 (2005).

\bibitem{VIC08} R. Vicente, L.~L. Gollo, C.~R. Mirasso, I. Fischer, and P. Gordon, Proc. Natl. Acad. Sci. U.S.A. {\bf 105},  17157  (2008).

\bibitem{WIL14} C. Wille, J. Lehnert, and E. Sch{\"o}ll, Phys. Rev. E {\bf 90}, 032908 (2014).

\bibitem{Mor13} I.-C. Morarescu, W. Michiels, M. Jungers, in Proc. Am. Control Conf., Washington (2013).

\bibitem{FFGHS} B. Fiedler, V. Flunkert, M. Georgi, P. H\"ovel, and E. Sch\"oll, Phys. Rev. Lett. {\bf 98}, 114101 (2007).

\bibitem{KBS11} Y.N. Kyrychko, K.B.  Blyuss, and E. Sch\"oll, Eur. Phys. J. B {\bf 84}, 307 (2011). 

\bibitem{KBS13} Y.N. Kyrychko, K.B.  Blyuss, and E. Sch\"oll, Phil. Trans. Roy. Soc. A {\bf 371}, 20120466 (2013).

\bibitem{PP} K. Pyragas and T. Pyragiene, Phys. Rev. E {\bf 78}, 046217 (2008).

\bibitem{Adimy} M. Adimy, F. Crauste, and S. Ruan, Nonl. Anal. RWA {\bf 6}, 651 (2005).

\bibitem{Barrio} M. Barrio, K. Burrage, A. Leier, and T. Tian, PLoS Comp. Biol. {\bf 2}, e117 (2006).

\bibitem{Blythe} S.P. Blythe, R.M. Nisbet, W.S.C. Gurney, and N. MacDonald, J. Math. Anal. Appl. {\bf 109}, 388 (1985).

\bibitem{CZ} K. Cooke and Z. Grossman, J. Math. Anal. Appl. {\bf 86}, 592 (1982).

\bibitem{Cush} J.M. Cushing, in {\it Mathematics of Biology}, edited by M. Iannelli (Springer, New York, 2011).

\bibitem{STE98a} G. St\'ep\'an, in {\it Dynamics and Chaos in Manufacturing Process}, edited by F.C. Moon (Wiley, New York, 1998).

\bibitem{Mit} J.E. Mittler, B. Sulzer, A.U. Neumann, and A.S. Perelson, Math. Biosci. {\bf 152}, 143 (1998).

\bibitem{SNA07} R. Sipahi, S.-I. Niculescu, and F.M. Atay, in Proc. 2007 Am. Control Conf., New York (2007).

\bibitem{Roe} O. Roesch, H. Roth, and S.-I. Niculescu, in Proc. IEEE Int. Conf. Mechatronics \& Automation, Niagara Falls (2005).

\bibitem{ZIL07} R. Zillmer, R. Livi, A. Politi, and A. Torcini, Phys.~Rev.~E {\bf 76},  046102 (2007).

\bibitem{PHT05} O.V. Popovych, C. Hauptmann, and P.A. Tass, Phys. Rev. Lett. {\bf 94}, 164102 (2005).

\bibitem{ZOU11} W. Zou, J. Lu, Y. Tang, C. Zhang, and J. Kurths, Phys. Rev. E {\bf 84}, 066208 (2011).

\end{thebibliography}
\end{document}